\documentclass{ws-ijmpd}
\usepackage[compress]{cite}
\usepackage{graphicx}
\usepackage{latexsym}
\usepackage{amssymb, amsmath, bm, physics,color, xcolor}
\usepackage[colorlinks,citecolor=blue,urlcolor=blue,linkcolor=blue]{hyperref}
\usepackage{orcidlink}

\begin{document}

\markboth{Kalita, Mukhopadhyay \& Govindarajan}
{White dwarfs in noncommutative geometry}

\catchline{}{}{}{}{}

\title{Significantly super-Chandrasekhar mass-limit of white dwarfs in noncommutative geometry}

\author{Surajit Kalita\orcidlink{0000-0002-3818-6037}}

\address{Department of Physics, Indian Institute of Science, Bangalore 560012, India\\
	surajitk@iisc.ac.in}

\author{Banibrata Mukhopadhyay\orcidlink{0000-0002-3020-9513}}

\address{Department of Physics, Indian Institute of Science, Bangalore 560012, India\\
	bm@iisc.ac.in}

\author{T. R. Govindarajan\orcidlink{0000-0002-8594-0194}}

\address{The Institute of Mathematical Sciences, Chennai 600113, India\\
	Chennai Mathematical Institute, Kelambakkam, Chennai 603103, India\\
	trg@imsc.res.in, trg@cmi.ac.in}

\maketitle

\begin{history}
\received{Day Month Year}
\revised{Day Month Year}
\end{history}

\begin{abstract}
	Chandrasekhar made the startling discovery about nine decades back that the mass of compact object white dwarf has a limiting value, once nuclear fusion reactions stop therein. This is the Chandrasekhar mass-limit, which is $\sim1.4M_\odot$ for a non-rotating non-magnetized white dwarf. On approaching this limiting mass, a white dwarf is believed to spark off with an explosion called type~Ia supernova, which is considered to be a standard candle. However, observations of several over-luminous, peculiar type~Ia supernovae indicate that the Chandrasekhar mass-limit to be significantly larger. By considering noncommutativity among the components of position and momentum variables, hence uncertainty in their measurements, at the quantum scales, we show that the mass of white dwarfs could be significantly super-Chandrasekhar and thereby arrive at a new mass-limit $\sim 2.6M_\odot$, explaining a possible origin of over-luminous peculiar type~Ia supernovae. The idea of noncommutativity, apart from the Heisenberg's uncertainty principle, is there for quite sometime, without any observational proof however. Our finding offers a plausible astrophysical evidence of noncommutativity, arguing for a possible second standard candle, which has many far-reaching implications.
	
\end{abstract}

\ccode{PACS Nos.: 02.40.Gh, 97.20.Rp, 04.20.-q, 97.10.Nf, 71.70.Di}

\section{Introduction}\label{Introduction}

Einstein's theory of general relativity (GR) and quantum mechanics are considered to be among the greatest discoveries in the twentieth century. GR is undoubtedly the most panoramic theory to explain the theory of gravity. It can easily explain a large number of phenomena where Newtonian gravity falls short. It also helps to understand the stability of Chandrasekhar's mass-limit for the white dwarf with a finite radius. The white dwarf is the end state of a star with mass $\lesssim 8 M_\odot$, where the inward gravitational force is balanced by the force due to outward electron degeneracy pressure arising from Fermi statistics. Moreover, if the white dwarf has a binary partner, it starts pulling matter out off the partner due to its high gravity, resulting in the increase in the mass of the white dwarf. When it gains a sufficient amount of matter, beyond a certain mass, known as the Chandrasekhar mass-limit (currently accepted value $\sim 1.4M_\odot$ \cite{1931ApJ....74...81C} for a carbon-oxygen non-magnetized and non-rotating white dwarf), this pressure/force balance is no longer sustained, and it sparks off to produce type Ia supernova (SNIa) \cite{choudhuri_2010}. The luminosity of SNIa is very important as it is used as one of the standard candles to measure the luminosity distance of various objects in cosmology.

However, recent observations have questioned the complete validity of GR near the compact objects. For example, in the past decade, a number of peculiar over-luminous SNeIa, viz. SN 2003fg, SN 2006gz, SN 2007if, SN 2009dc \cite{2006Natur.443..308H,2010ApJ...713.1073S}, etc. have been observed, which were inferred to be originating from white dwarfs of super-Chandrasekhar mass as high as $2.8M_\odot$. In this scenario, the Chandrasekhar mass-limit is well violated. Different theories and models have been proposed to explain this class of the white dwarfs. Our group started exploring the significant violation of the Chandrasekhar mass-limit based on the effect of magnetic fields \cite{2012MPLA...2750084K,2012PhRvD..86d2001D}. Subsequently, there is enormous interest in re-exploring the Chandrasekhar mass-limit by introducing various new physical effects in white dwarfs. Some such physics are (1) effects of the strong magnetic field leading to significantly super-Chandrasekhar mass: quantum, through Landau orbital effects above the Schwinger limit $4.414\times 10^{13}$ G, which affects the equation of state (EoS) \cite{2013PhRvL.110g1102D}, and classical: through the field pressure affecting the macroscopic structural properties \cite{2014JCAP...06..050D,2015JCAP...05..016D,2015MNRAS.454..752S,2019MNRAS.490.2692K,2020IAUS..357...79K}; (2) modified gravity effect, leading to significantly sub- and super-Chandrasekhar mass-limits \cite{2015JCAP...05..045D,2017EPJC...77..871C,2018JCAP...09..007K}; (3) ungravity effect \cite{2016PhRvD..93j4046B}; (4) consequence of total lepton number violation in magnetized white dwarfs \cite{2015NuPhA.937...17B}; (5) charged white dwarfs leading to super-Chandrasekhar mass \cite{2014PhRvD..89j4043L}; (6) generalized Heisenberg uncertainty principle \cite{2018JCAP...09..015O}; (7) effects of momentum-momentum noncommutativity in the white dwarf matter and hence the equation of state, leading to super-Chandrasekhar mass-limit \cite{2019PhLB..79734859P}; and many more. It is, however, important to emphasize that the super-Chandrasekhar white dwarfs are predicted from indirect observations. In other words, to date, none of the super-Chandrasekhar white dwarfs have been detected directly. Therefore, nobody can rule out any of these theories, or single out the exact theory of their existence. Perhaps time will determine which one is the key theory. In the premise of astrophysical phenomenology, the said anomalous brightness has been attempted to explain by plausible non-ideal properties, such as departure from spherical symmetry and collision of the debris with a binary companion or other surrounding matter, off-center or otherwise asymmetric thermonuclear burning and possible instabilities in the flow due to uncertain opacities of the site, etc. \cite{2004ApJ...615..444S,2006MNRAS.373..263M,2009ApJ...702..686C,2019MNRAS.483..263W,2010Natur.463...61P,2014NewAR..62...15R,2018PhR...736....1L,2019arXiv191201550S}. However, none of the ideas could explain the existence of progenitor white dwarf mass as high as $2.8M_\odot$, as inferred from observation mentioned above.

The plan in the present work is to introduce position-position and momentum-momentum noncommutativities, in addition to the Heisenberg algebra, and analyze their effects on white dwarfs. Many researchers earlier used the idea of noncommutativity to explain the physics of various systems. Madore \cite{1992CQGra...9...69M} introduced the idea of a fuzzy sphere, which later has been used to describe Landau levels \cite{2003JHEP...08..057L,2014JPhA...47R5203C,2015JPhA...48C5401A,2015PhRvD..92l5013S}. Moreover, in the presence of noncommutativity, one can show that the spacetime metric alters \cite{2006LNP...698...97N}. As a result, in NC spacetime, the event horizon shifts, and the essential singularity at the center of a black hole can be removed \cite{2017EPJC...77..577K}, which further results in changing various other properties of black holes, such as the Hawking temperature \cite{2018EPJP..133..421F}. Eventually, many researchers have used different forms of NC geometry to describe various other problems of physics related to fundamental length scale, Berry curvature, Landau levels, gamma-ray burst, etc. \cite{1999JHEP...09..032S,2000IJMPA..15.4301A,2001PhLB..510..255A,2002PhRvL..88s0403M,2002Natur.418...34A,2012PhRvL.109r1602S}. However, unfortunately, there is no direct way to confirm the natural evidence of such noncommutativity and hence it still remains as a hypothesis. Nevertheless, our observable universe abides by position-position and momentum-momentum commutative rules, which implies that two position coordinates and two momentum coordinates can be measured simultaneously. However, at a microscopic length scale (and/or at a very high energy regime), the position and corresponding conjugate momentum follow the Heisenberg's uncertainty principle (i.e., they do not commute). The additional proposals are that position-position noncommutativity arises \cite{2001PhLB..510..255A,2002PhRvL..88s0403M,2006PhLB..632..547N,2017EPJC...77..577K,2018EPJP..133..421F} at the very high energy regime, e.g. at the Planck scale. On the other hand, the density and the corresponding energy scale of white dwarfs are significantly lower than those at the Planck scale and, hence, any implementation of the position-position noncommutativity in the white dwarf matter may be considered at a faith of strong hypothesis. Moreover, the Chandrasekhar mass-limit arises from the interplay of pressures due to fermionic statistics and gravitational attraction. One of the important outcomes of NC geometry is that the statistics of particles gets modified due to the starproduct \cite{2007PhRvD..75d5009B,2006JPhA...39.9557C}. Effectively a fermion behaves less of a fermion and, hence, the repulsive pressure among the `dressed particles' is reduced, and/or allowing collapse continuing till smaller radius, with more mass accumulated to attain stability. Therefore, although the scale of NC geometry in quantum spacetime, namely the Planck length, is too small, the coherent effect of large density of white dwarf can enhance the effective NC scale to a larger value. This will be argued with the realistic densities of white dwarfs taken into account.

One way of interpreting this noncommutativity is the existence of spacetime magnetic field, almost equivalent to Landau quantization. In the case of Landau quantization, position coordinates perpendicular to the direction of the magnetic field become NC in the presence of an external magnetic field and, hence, the corresponding generalized momentum components also become NC. It is a single parameter, i.e., field strength, which controls the noncommutativity of position and momentum coordinates. Now, the hypothesis in NC geometry is that there is an effective inherent magnetic field in the spacetime itself at the microscopic level in place of an external field. If so, at which length scale such a field, equivalent to external field producing Landau orbitals, becomes significant is a big question mark. However, if noncommutativity is present, with the analogy of external field effect, a single parameter should control both the position and momentum noncommutativities apart from the Heisenberg's uncertainty principle. Note that the position-position noncommutativity is more fundamental to describe the NC universe, and the momentum-momentum noncommutativity may arise as a consequence of it. Indeed it is a matter of the fact that the curvature in position space leads to the noncommutativity in the conjugate momentum variables. Interestingly, in the energy dispersion relation, only the momentum-momentum noncommutativity parameter appears explicitly, hence mathematically speaking, whether position coordinates commute or not, that does not matter. However, given the lack of information at this scale, it is not generic to assume only momentum-momentum NC relation.

Recently, momentum-momentum noncommutativity has been hypothesized, keeping however, position-position commutativity intact \cite{2019PhLB..79734859P}, which has also been argued to be a dynamical noncommutativity. However, as discussed above, NC geometry primarily offers a position-space noncommutativity that may lead to a momentum-momentum noncommutativity similar to the effect of the external magnetic field in Landau quantization. Moreover, the said work \cite{2019PhLB..79734859P} argues the mass-limit of the white dwarf to be $4.68M_\odot$, which seems to be unrealistic from the observations, at least in the present formalism. One may assume that at a high central density $\sim 10^{10}$ g cm$^{-3}$, such possibility may arise if the noncommutativity effect is triggered at such density/energy. Indeed, it was shown earlier that such a high mass-limit of white dwarfs is not ruled out in the presence of a magnetic field \cite{2013IJMPD..2242004D}. Nevertheless, the idea those authors proposed was that the field pressure (acting outward) and field density, as well as tension (may act inward), would help to gravitate the star and hence to hold extra mass \cite{2013IJMPD..2242004D}. It is a macroscopic effect, unlike what is proposed based on the momentum-momentum noncommutativity by the authors \cite{2019PhLB..79734859P}, which is purely microscopic. However, at a low density, e.g., near the surface of the white dwarf or the low central density white dwarfs, which are often evident observationally \cite{2018MNRAS.480.4505J}, the same noncommutativity effect should be inactive, and the known observations should be explained. This should be considered to impose constraints on the NC parameter. However, the authors \cite{2019PhLB..79734859P} ignored this crucial physical fact to explain the low central density white dwarfs. Any successful theory should explain the entire set of observations (for the present purpose, both high and low density white dwarfs). In fact, the authors \cite{2019PhLB..79734859P} neither considered general relativistic treatment nor thoroughly explored the mass-radius relation, which is inevitable in order to assess the reality of the results. The former is essential to establish that the limiting mass of the white dwarfs is at a finite radius and finite density, thereby providing an instability point in the mass-radius curve, which is missing in the Newtonian treatment. The latter will assure the consistency of the results at the low density regime by fixing the NC parameter at the center.

The present work reasonably overcomes both the shortcomings along with the previous misleading results \cite{2019PhLB..79734859P} and shows that a single NC parameter can control both the position- and momentum-space noncommutativities, similar to the magnetic field, which affects the underlying microphysics and thereby the stellar structure. Furthermore, we constrain the NC parameter appropriately to reveal the known low density results by the same formalism. We show that with proper constraints, the mass-limit is actually $\sim 2.6M_\odot$, which has already been reported from observations. The plan of the paper is as follows. In Section \ref{Formalism of noncommutativity}, we discuss the formalism of the noncommutativity based on which we calculate further the energy spectrum or the dispersion relation in Section \ref{Energy spectrum}. Eventually, we use this relation to obtain the degenerate equation of state for the electrons in Section \ref{Degenerate equation of state}. In this Section, we also discuss the noncommutativity parameter, which defines the energy spectrum. Further, in Section \ref{Mass-radius relation and limiting mass of white dwarfs}, we obtain the mass-radius relation along with the new mass-limit of the white dwarf in the presence of noncommutativity before we conclude in Section \ref{Conclusions}.


\section{Noncommutativity model}\label{Formalism  of noncommutativity}

The NC algebra on the two dimensional plane has a direct link with the Landau quantization in the presence of magnetic field. The Landau problem is perhaps the simplest example of a system exhibiting spatial or momentum noncommutativity. By explicit investigation, one can find that the projection of coordinates to the Landau levels results in a NC algebra between the position coordinates of a particle in a two-dimensional plane. Inspired by the consequence of Landau effect, we propose our NC algebra. For relativistic electrons of mass $m_e$ moving in a three-dimensional space where $x-$ and $y-$coordinates constitute a NC Moyal plane \cite{1949PCPS...45...99M,1946Phy....12..405G}, our proposed NC Heisenberg algebra (NCHA), satisfied by the operators $\left(\hat{x}_i , \hat{p}_i\right)$, goes as follows
\begin{eqnarray}\label{NCHA}
&&\comm{\hat{p}_{j}}{\hat{p}_{k}}=i\theta \epsilon_{jk},~
\comm{\hat{x}_{j}}{\hat{p}_{k}} = i\hbar\delta_{jk},~
\comm{\hat{x}_{j}}{\hat{x}_{k}}=i\frac{\theta\eta^2}{4\hbar^2} \epsilon_{jk},\nonumber \\ && 
\comm{\hat{x}_j}{\hat{z}} = \comm{\hat{p}_j}{\hat{p}_z}=0,~~ \comm{\hat{z}}{\hat{p}_z} = i\hbar, \nonumber \\&&
\comm{\hat{x}_j}{\hat{p}_z} = 0, ~~ \comm{\hat{p}_j}{\hat{z}} = 0,
\end{eqnarray}
for $j, k = 1,2$ where subscripts $1$ and $2$ respectively imply $x-$ and $y-$components of respective variables. Here $\theta$ is the NC parameter, $\eta$ an arbitrary constant which takes care of the dimension of the position-position noncommutativity, $\hbar$ the reduced Planck constant, $\delta_{jk}$ the Kronecker delta tensor and $\epsilon_{jk}$ the antisymmetric Levi-Civita tensor. Since only the $x-y$ plane is noncommutative, the motion along $z$-direction is free and commutes with the $x$ and $y$ coordinates. This apparently violates the $\mathtt{SO}(3)$ symmetry. This can however be restored for a system with $\theta = 0$ as it recovers the Heisenberg algebra. Nevertheless, for $\theta\neq 0$, which is the present interest, the breaking spherical symmetry is in microscopic scale, which does not necessarily influence macroscopic structure of the star, which may remain spherical unless other macroscopic effects like rotation, magnetic fields are introduced. However, in future, the plan is to explore the introduction three-dimensional NC effect, e.g. fuzzy sphere \cite{1992CQGra...9...69M,2015JPhA...48C5401A}, in white dwarf. It is important to remember that since we deal with the compact objects, any NC algebra, whether two- or three-dimensional, is required to achieve the fact that there is a space dependent noncommutativity which is maximum at the core and goes to zero at the surface in a phenomenological way.

The standard approach in the literature to deal with such problems is to form an equivalent commutative description of the NC theory by employing some non-canonical transformation, the so-called Bopp shift, which relates the NC operators $\hat{x}_{j}$, $\hat{p}_{j}$ following equation (\ref{NCHA}) to ordinary commutative operators $x_{j}$, $p_{j}$, satisfying the usual Heisenberg algebra 
\begin{eqnarray}
\comm{x_{j}}{p_{k}}=i\hbar \delta_{jk}, \quad \comm{x}{y}=0= \comm{p_{x}}{p_{y}}.
\label{HA}
\end{eqnarray}
In our subsequent discussion, we denote NC operators with the hat notation and commutative operators without hat, and to satisfy the above NC algebra, we use the following generalized Bopp-shift transformations which is given by
\begin{eqnarray}
\label{e74}
\hat{p}_{j}  =  p_{j}+\frac{\theta}{2\hbar}\epsilon_{jk}x_{k}, ~~~
\hat{x}_{j}  =  x_{j}+\frac{\eta}{2\hbar}\hat{p}_{j}.
\end{eqnarray}

If the total Hamiltonian of the system is $\hat{H}$, the Dirac equation for an electron moving in the NC plane satisfying the NCHA reads
\begin{equation}
\hat{H} \psi =i\hbar\pdv{\psi}{t} = E \psi,
\end{equation}
where $\psi$ is a two-component spinor of components $\phi$ and $\chi$ with the Dirac Hamiltonian is given by 
\begin{equation}
\hat{H}=\bm{\alpha}\vdot\bm{\hat{p}}c+ \beta m_e c^2,
\end{equation}
where $c$ is the speed of light and $\bm{\alpha}$ and $\beta$ have their usual meaning. The above gives a pair of equations
\begin{eqnarray}
\left(E-m_ec^2\right)\phi=\bm{\sigma}\vdot\bm{\hat{p}}c\, \chi \,\,
{\rm and}\,\,
\left(E+m_ec^2\right)\chi=\bm{\sigma}\vdot\bm{\hat{p}}c\, \phi,
\end{eqnarray}
where $\bm{\sigma}$ is the Pauli matrix in vector form. On combining them, we obtain
\begin{eqnarray}\label{energy}
E^2-m_e^2c^4 
&=& \left(\bm{\sigma}\vdot\bm{\hat{p}}\right)^2c^2 \nonumber \\
&=& \bm{\hat{p}}^2c^2 + i \bm{\sigma}\vdot\left(\bm{\hat{p}}\times\bm{\hat{p}}\right)c^2 \nonumber \\
&=& \Big[p_z^2 -\sigma_z \theta+ \left(p_x^2 + p_y^2\right)+ \frac{\theta^2}{4\hbar^2}\left(x^2+y^2\right) + \frac{\theta}{\hbar}\left(y p_x -x p_y\right)\Big]c^2.
\label{eigen}
\end{eqnarray}
Therefore, we obtain an equivalent commutative Hamiltonian in terms of the commutative variables (quantum mechanical operators) which describes the original system defined over the NC plane.


\section{Energy dispersion relation}\label{Energy spectrum}
To compute the spectrum of a charged particle in such a NC spacetime, first of all we need to construct the ladder operators which will diagonalize the following part of right hand side of equation (\ref{eigen}), given by
\begin{equation}\label{diag}
\hat{H}^{\prime}= \left[\left(p_x^2 + p_y^2\right)+ \frac{\theta^2}{4\hbar^2}\left(x^2+y^2\right) + \frac{\theta}{\hbar}\left(y p_x -x p_y\right)  \right]c^2.
\end{equation}
The ladder operators involving the commutative phase-space variables (operators) $x$, $y$, $p_{x}$, $p_{y}$, given by 
\begin{eqnarray}\label{e30a}
a_{j} = \frac{1}{\sqrt{\theta}} \left(p_j-i\frac{\theta}{2\hbar}x_j\right), \quad
a_{j}^{\dagger} = \frac{1}{\sqrt{\theta}} \left(p_j+i\frac{\theta}{2\hbar}x_j\right),\nonumber
\end{eqnarray}
satisfy the commutation relations 
\begin{eqnarray}
\label{e30}
\comm{a_{1}}{a_{1}^{\dagger}}=1=\comm{a_{2}}{a_{2}^{\dagger}}.
\end{eqnarray}
Further defining a pair of operators
\begin{eqnarray}
\label{lad}
\hat{b}_{1}=\frac{a_{1}+ia_{2}}{\sqrt{2}},\quad \hat{b}_{2}=\frac{a_{1}-ia_{2}}{\sqrt{2}},
\end{eqnarray}
which satisfy the commutation relations
\begin{eqnarray}
\label{e32a}
\comm{\hat{b}_{1}}{\hat{b}_{1}^{\dagger}} =1 =\comm{\hat{b}_{2}}{\hat{b}_{2}^{\dagger}},
\end{eqnarray}
the Hamiltonian given by equation (\ref{diag}) can be recast in the diagonal form as
\begin{eqnarray}
\label{diagfinal}
\hat{H}^{\prime}=\theta \left(2\hat{b}_{1}^{\dagger} \hat{b}_{1} +1\right)c^2.
\end{eqnarray}
Therefore, combining (\ref{energy}) and (\ref{diagfinal}), the total energy of the system is given by
\begin{equation}\label{energy_spectrum}
E^2(\nu)=p_{z}^2 c^2 + m_e^2c^4 +2 \nu \theta c^2,
\end{equation}
where for spin-up $\left(s=1/2\right)$, $\nu=n_1$ and for spin-down $\left(s=-1/2\right)$, $\nu=n_1+1$, when $n_1$ is the eigenvalue of the number operator $\hat{b}_{1}^{\dagger} \hat{b}_{1}$. It is of course important to note that $\theta \to 0$ does not automatically lead to the commutative results as the creation and annihilation operators $\left(a,~ a^{\dagger}\right)$ are inversely proportional to $\theta$. Hence, as $\theta \to 0$, $a$ and $a^{\dagger}$ behave `singular' hence $\nu$ turns out to be huge (or divergent). However, their product $\left(\theta\nu\right)$ turns out to be finite, which, for instance, appears in the energy expression given by equation \eqref{energy_spectrum}, not $\theta$ and $\nu$ independently. This is exactly the case for Landau quantization in the presence of external magnetic field, where the magnetic field plays the same role of $\theta$ here. Indeed in the quantum mechanics based on Heisenberg algebra, $\hbar$  associated with the phase-space noncommutativity cannot be chosen to be zero just to reproduce classical results.


\section{Equation of state of matter}\label{Degenerate equation of state}
Although the EoS, for the above dispersion relation, has already been found earlier \cite{2012PhRvD..86d2001D,2019PhLB..79734859P}, just for completeness, in this section, we briefly discuss about it. Using equation \eqref{energy_spectrum}, the Fermi energy $E_F$ of electrons for the $\nu$th level is given by
\begin{equation}
E_F^2(\nu)=p_{zF}^2(\nu) c^2 + m_e^2c^4 +2 \nu \theta c^2,
\end{equation}
where $p_{zF}$ is the Fermi momentum of the electrons. In dimensionless form, it can be recast as follows
\begin{equation}\label{nu}
\epsilon_F^2(\nu)=x_F^2(\nu) + 1 +2 \nu \theta_D,
\end{equation}
where $\theta_D=\frac{\theta}{m_e^2c^2}$, $\epsilon_F=\frac{E_F}{m_e c^2}$ and $x_F(\nu)=\frac{p_{zF}(\nu)}{m_e c}$.

Due to the quantization of the energy levels in the $x-y$ plane, the modified density of state becomes $\left(4\pi \theta/h^3\right)\dd{p_z}$. Hence the electron number density and electron energy density at zero temperature are respectively given by
\begin{align}\label{density}
n_e&=\sum_{\nu=0}^{\nu_{max}} \frac{4\pi m_e^3 c^3 \theta_D}{h^3}g_\nu x_F(\nu),\\
u&=\frac{4\pi m_e^3 c^3 \theta_D}{h^3}\sum_{\nu=0}^{\nu_{max}} g_\nu \int_{0}^{x_F} E(\nu) \dd{x(\nu)},
\end{align}
where $g_\nu$ is the degeneracy such that $g_\nu=1$ for $\nu=0$ and $g_\nu=2$ for $\nu \ge 1$, which is taken to be the nearest lowest integer of $(\epsilon_F^2-1)/2\theta_D$ for every $\epsilon_F$ and $\theta_D$. Therefore the pressure of the Fermi gas for the electrons is given by
\begin{align}\label{pressure}
P=&n_e E_{F}- u \nonumber \\
=&\sum_{\nu=0}^{\nu_{max}} \frac{2\pi m_e^4 c^5 \theta_D }{h^3}g_\nu \bigg[ \epsilon_{F} x_F(\nu) - (1+2\nu\theta_D) \log \left(\frac{\epsilon_{F}+ x_F(\nu)}{\sqrt{1+2\nu\theta_D}}\right)\bigg],
\end{align}
and the mass density is given by
\begin{equation}\label{rho number density relation}
\rho=\mu_e m_p n_e,
\end{equation}
where $\mu_e$ is the mean molecular weight per electron and $m_p$ is the mass of a proton.

Now we assume that all the electrons are filled in the lowest Landau level, which implies that $\nu=0$. The validity and condition of this assumption are given below. For $\nu=0$,
\begin{equation}\label{ground_lavel}
\rho =Qx_F(0),
\end{equation}
and the EoS given by equation \eqref{pressure} reduces to
\begin{equation}\label{non_comm_EoS}
P=\frac{h^3}{8\pi \mu_e^2 m_p^2 m_e^2 c \theta_D } \left[ \rho \sqrt{\rho^2+ Q^2} -Q^2 \log \frac{\rho+ \sqrt{\rho^2+ Q^2}}{Q} \right],
\end{equation}
where
\begin{equation}\label{Q_eta}
Q=\frac{4\pi\mu_e m_pm_e^3c^3}{h^3}\theta_D.
\end{equation}

Let us now look at the asymptotic behavior of this EoS. For $x_F(0)\gg 1$, which corresponds to $\rho^2 \gg Q^2$, EoS further reduces to the following simpler polytropic form
\begin{equation}
P=\frac{h^3}{8\pi \mu_e^2 m_p^2 m_e^2 c \theta_D }\rho^2=K_{nc} \rho^2=K_{nc}\rho^{1+1/n},
\end{equation}
with the polytropic index $n=1$.

However, for the present case, also $x_F^2=\epsilon_F^2-1>0$ which implies that 
\begin{equation}\label{nu1}
\epsilon_F^2=2\nu_1\theta_D+1,
\end{equation}
where $0\lesssim \nu_1<1$, particularly at the center and for ground level (equivalent to the lowest Landau level for the magnetic case) $\sqrt{\epsilon_F^2(0)-1}= x_F(0) = \sqrt{2\nu_1\theta_D}$, which implies from equation \eqref{ground_lavel}
\begin{equation}\label{rho2}
\rho=\frac{4\pi\mu_e m_p m_e^3 c^3}{h^3}\theta_D^{3/2}\sqrt{2\nu_1},
\end{equation}
when $\nu_1$ can have any value below unity for all electrons to be in the ground level. Rewriting $\theta_D$ in terms of $\rho$ and substituting it in equations \eqref{non_comm_EoS} and \eqref{Q_eta}, we have the EoS of the degenerate matter of the white dwarf, which is given by
\begin{equation}\label{NCM_EoS}
\begin{aligned}
P &= \left(\frac{A\nu_1}{\rho^2}\right)^{1/3} \left[ \rho \sqrt{\rho^2+ Q^2} -Q^2 \log \frac{\rho+ \sqrt{\rho^2+ Q^2}}{Q} \right],\\
Q &= \left(\frac{B}{\nu_1}\right)^{1/3}\rho^{2/3},
\end{aligned}
\end{equation}
with $$A=\frac{h^3 c^3}{16\pi \mu_e^4 m_p^4}, ~~~ B=\frac{2\pi\mu_e m_p m_e^3 c^3}{h^3}.$$ At large density limit, the above EoS given by equation \eqref{NCM_EoS} reduces to
\begin{equation}\label{relativistic_EoS}
P = \left(\frac{h^3 c^3 (2\nu_1)}{32\pi \mu_e^4 m_p^4}\right)^{1/3}\rho^{4/3} = K_{ncm}\nu_1^{1/3}\rho^{4/3}.
\end{equation}
This is the highly relativistic EoS of degenerate electron gas, which we further use to compute the limiting mass of the white dwarf in the presence of noncommutativity. Here $\nu_1$ defines how much a Landau level (here the ground level) fills in.


\subsection{Constraining noncommutativity parameter} \label{Constraining noncommutativity parameter}

If we consider the density at center $\rho_c=2\times 10^{10}/V$ g cm$^{-3}$,
from equation (\ref{rho2}), 
we obtain the central $\theta_D$ of the star
\begin{align}
\theta_D &= \left(\frac{2\times 10^{10}h^3}{4\pi\mu_e m_p m_e^3 c^3\sqrt{2\nu_1}V}\right)^{2/3} \approx \frac{456}{(V\mu_e)^{2/3}\nu_1^{1/3}}.
\label{theta}
\end{align}
The parameter $V$ is introduced to determine the deviation of $\rho_c$ from a typical central density giving rise to the Chandrasekhar mass-limit of a nonmagnetized, nonrotating white dwarf. For $V=1$ of a carbon-oxygen white dwarf when $\mu_e=2$, $\theta_D\sim 287.3$ at the center from equation (\ref{theta}). This clearly confirms from equation \eqref{rho2} that from the center to surface, average distance between the electrons increases as the density decreases and, hence, the position and momentum spaces, both tend to become commutative.

Let us now pay attention to the values of the noncommutativity parameters. As discussed above, at the center of a white dwarf with $\rho_c=2\times10^{10}$ g cm$^{-3}$, we have $\theta_D\sim287.3$, which implies $\theta = \theta_D m_e^2c^2 \sim 2.1\times10^{-31}$ g$^2$ cm$^2$/s$^2$. However, the Planck energy is $\sim 2\times10^{16}$ ergs, which implies the Planck momentum to be $\sim 6.5\times10^5$ g cm/s. It clearly indicates that our regime of interest is very far from the Planck scale, but it still does not exactly follow the commutative algebra. This small deviation from the commutative algebra is enough to show the violation of the Chandrasekhar mass-limit. It is also clear from equation \eqref{rho2} that in case of neutron stars whose $\rho_c$ is much higher as compared to the white dwarf, the effect of noncommutativity will be much more significant. In both the cases, as density decreases from the center to surface, it is expected that the physics would be dominant by the usual commutative algebra, which is also evident from equation \eqref{rho2}. At the surface, density decreases practically to zero, which indicates no noncommutativity and the usual commutative algebra is restored. Moreover, the constant $\eta$, introduced for the position-position noncommutativity in equation \eqref{NCHA}, represents the typical length scale of the system at which the position-position noncommutativity is significant. The value of $\theta/4\hbar^2$ is $\sim 4.8\times10^{22}$ cm$^{-2}$ for a white dwarf with $\rho_c=2\times10^{10}$ g cm$^{-3}$ and is $\sim 2.2\times10^{25}$ cm$^{-2}$ for a neutron star with $\rho_c=2\times10^{14}$ g cm$^{-3}$. From equation \eqref{NCHA}, the value of $\eta$ has to be chosen in such a way that the position noncommutativity is much smaller as compared to the Planck scale.


\subsection{Scale of noncommutativity in white dwarfs}
The scale of NC parameter emerges from inherent coarse grained/foamy structure of the spacetime whose fundamental length is obtained as the Planck length \cite{2011PhRvD..83h4003C}. However, quantum physics of matter through coherent effects brings additional structures to qualitatively enhance this length. This kind of expectation of quantum gravity effects, reflecting in stellar objects under extreme conditions, have been studied earlier. For example, the uncertainty in measurement of particular distance `$L$' was anticipated to be $(L~L_P^2)^\frac{1}{3}$, where $L_P$ being the Planck length. Such an argument is motivated by the arguments of Salecker and Wigner \cite{1958PhRv..109..571S}.

The following arguments are given by Ng \cite{2003MPLA...18.1073N, 2019Entrp..21.1035N}. Let us assume the distance between two points $A$ and $B$ to be $L$. We can consider a clock of mass $m$ described by a wavepacket with $\delta$ as its spread. It sends a light signal from $A$ to $B$ which gets reflected and returns back to $A$ at a time $\frac{2L}{c}$. The spread now results in a new spread $\delta + \frac{2\hbar~L}{mc\delta}$. In addition, GR provides another uncertainty namely the clock must have size $\delta \geq \frac{Gm}{c^2}$. These two arguments together yield $$\delta \geq \left(L~L_P^2 \right)^\frac{1}{3}.$$ Although it is heuristic, it combines both quantum mechanical and gravity effects for massive compact objects with large densities where interparticle distances are comparable or even much less than the Compton wavelength.

We can look for measuring the distance between atoms in white dwarfs. For a white dwarf with density $\rho\sim 10^7$ g cm$^{-3}$, the inter-atomic distance turns out to be $L\sim (10^7/m_p)^{-1/3}$. This works out to be $L\sim 10^{-10}$ cm, which is of the order of the Compton wavelength of electrons. Now, for higher densities such as $\rho \gtrsim 10^{10}$ g cm$^{-3}$, which is for typical super-Chandrasekhar white dwarfs, we obtain $L \sim 10^{-12}$ cm, which is two orders of magnitude less than the Compton wavelength. The nature of statistics changing due to NC geometry is expected to play a crucial role. If we use the scale provided by the previous argument, we find $\delta\sim 10^{-26}$ cm, which is expected to be increased further due to many particle effects in a white dwarf. For usual white dwarfs, the inter-electron distance will be more than the Compton wavelength, thereby conventional Fermi statistics will be sufficient for getting the equation of state. A word of caution may be in order here however. The scale of uncertainty in no way provides the details of NC geometry. Therefore, we assume simple generalization of Moyal plane.


\section{Obtaining mass-limit}\label{Mass-radius relation and limiting mass of white dwarfs}

To obtain the interior solution of any star, one needs to solve simultaneously the mass and momentum balance equations (together known as Tolman-Oppenheimer-Volkoff equations or in short TOV equations in GR) with appropriate boundary conditions. In GR, for static, non-magnetized fluid, the energy-momentum tensor is given by $$T^\mu_\nu = \text{diag}(-\rho c^2, P, P, P),$$ with $\rho$ being the density and $P$ being the pressure of the fluid at an arbitrary $r$. Using the conservation of the energy-momentum tensor $\nabla_\mu T^\mu_\nu =0$, the TOV equations in GR are given by \cite{2009igr..book.....R}
\begin{equation}\label{TOV}
\begin{aligned}
\dv{M}{r} &= 4\pi r^2\rho,\\
\dv{P}{r} &= -\frac{(P+\rho c^2)}{1-\frac{2GM}{c^2r}}\left[\frac{G}{r^2}\left(\frac{4\pi r^3P}{c^4}+\frac{M}{c^2}\right)\right],
\end{aligned}
\end{equation}
where $M$ is the mass of the star inside the volume of radius $r$ and $G$ the gravitational constant. Now we need to solve these two differential equations simultaneously to obtain the interior solution of the star with the EoS given by equation \eqref{NCM_EoS}. The boundary conditions at the center of the star are $M (r =0) = 0$ and $\rho(r = 0) = \rho_c$. On the other hand, at the surface of the star, we have $\rho(r = R_*) = 0$ and $M (r = R_*) = M_*$, the total mass of the star, where $R_*$ is the radius of the star. We consider the central density $\rho_c$ of the white dwarfs to be varied from $2\times 10^5$ g cm$^{-3}$ to $10^{11}$ g cm$^{-3}$ to avoid the neutron drip.

\begin{figure*}
	\centering
	\includegraphics[scale=0.7]{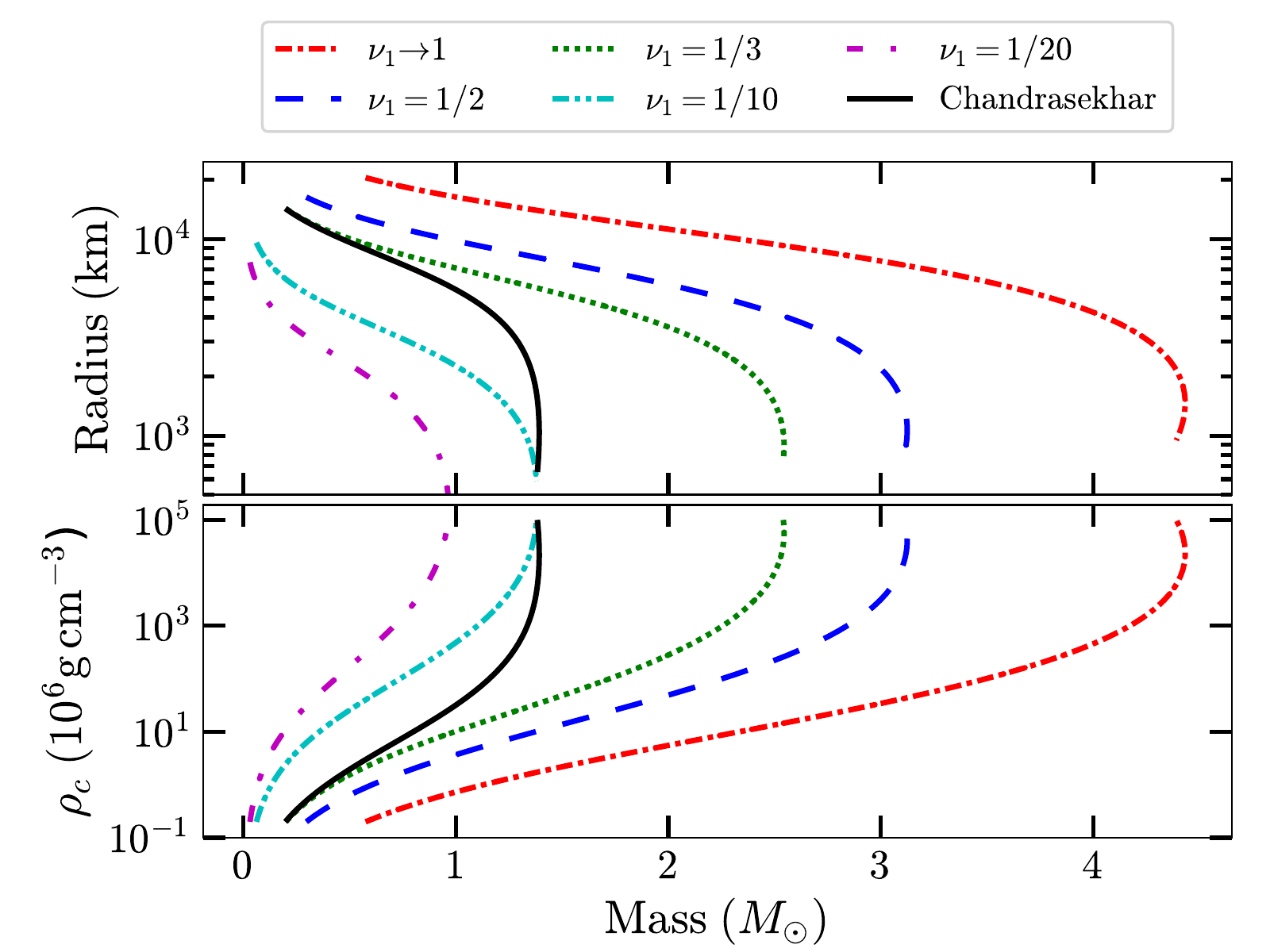}
	\caption{Upper panel: The mass-radius relation, Lower panel: The variation of central density with mass of the white dwarf. Black solid line represents the mass-radius relation found by Chandrasekhar, whereas other lines represent the same in presence of noncommutativity.}
	\label{Fig: mass_radius}
\end{figure*}

Figure \ref{Fig: mass_radius} shows the mass-radius relation as well as the variation of the central density with respect to the mass of white dwarfs for various values of $\nu_1\lesssim 1$ along-with the same for Chandrasekhar's EoS. It is clearly seen that the curves attain a peak and then turn back above a certain $\rho_c$, which means that the curves become unstable after that point, indicating a limiting mass of the white dwarf for each values of $\nu_1$. However, for $\nu_1=1/3$, the curve matches with the curve for the Chandrasekhar's EoS at low density, which seems to be physical, as most of the white dwarfs those have been observed are bigger in size \cite{2018MNRAS.480.4505J} (which correspond to the low density white dwarfs), following Chandrasekhar's result. Hence the mass-radius relation should not be violated from the original mass-radius relation at low $\rho_c$, indicating very negligible effect of noncommutativity at the lower density. However, at a higher value of $\rho_c$, the mass-radius relation can be significantly differed from the original one due to the presence of noncommutativity, and, as a result, the limiting mass of the white dwarf turns out to be super-Chandrasekhar with $M_*\sim2.6M_\odot$ for $\nu_1 = 1/3$, implying the violation of the Chandrasekhar mass-limit in the presence of NC geometry.

Now we are going to discuss about the limiting mass of the white dwarf using Lane-Emden formalism \cite{1964ApJ...140..434T,1973ApJ...183..637B}. The hydrostatic balance equations are given by the relations \eqref{TOV} and the limiting form of EoS for the degenerate electron gas is given by $$P=K\rho^{1+1/n}.$$ Defining the dimensionless density $\Theta$ as $\rho = \rho_c \Theta^n$, the above equation can be written as 
\begin{equation}
\frac{P}{\rho c^2} = \frac{K}{c^2}\rho^{1/n} = \sigma\Theta,
\end{equation}
with $\sigma = \rho_c^{1/n} K/c^2$. Now, let us introduce the dimensionless distance $\xi$ and dimensionless mass $\mu$ such that $r=a\xi$ and $\mu(\xi) = \left(1/4\pi\rho_c a^3\right) M(r)$, where
\begin{equation}
a = \left[\frac{(n+1)K \rho_c^\frac{1-n}{n}}{4\pi G}\right]^{1/2}.
\end{equation}
Using these dimensionless variables, equations \eqref{TOV} can be recast as
\begin{align}
\dv{\mu}{\xi} &= \xi^2\Theta^n, \\
-\xi^2\dv{\Theta}{\xi} &= \frac{\left(\mu+\sigma\Theta\xi \dv*{\mu}{\xi}\right) \left(1+\sigma\Theta\right)}{1-2\sigma\left(n+1\right)\mu/\xi}.
\end{align}
These set of equations are known as the Lane-Emden equation in GR. Note that, for $\sigma\to 0$, the above equations reduce to the Newtonian Lane-Emden equations. Now, the mass of the white dwarf is given by
\begin{align*}
M_* &= \int_0^R 4\pi r^2 \rho \dd{r} = 4\pi a^3 \rho_c \int_0^{\xi_1}\xi^2\Theta^n \dd{\xi},
\end{align*}
where $\xi_1$ is defined as $R_*=a\xi_1$. From equation \eqref{relativistic_EoS}, it is observed that for the present case, $n=3$ and $K=K_{ncm}\nu^{1/3}$. Hence the mass of the white dwarf is given by
\begin{align*}
M_* &= 4\pi a^3 \rho_c \int_0^{\xi_1}\xi^2\Theta^3 \dd{\xi} \\
&= 4\pi \left[\frac{K_{ncm}^3\nu_1}{(\pi G)^3}\right]^{1/2} \int_0^{\xi_1}\xi^2\Theta^3 \dd{\xi} \\
&= 4\pi \left[\frac{K_{ncm}^3\nu_1}{(\pi G)^3}\right]^{1/2} \mu(\xi_1),\\
&\simeq 4.5 \sqrt{\nu_1}\; M_\odot,
\end{align*}
which is the limiting mass of the white dwarf. For $\nu_1 = 1/3$, the limiting mass of the white dwarf turns out to be $\sim 2.6M_\odot$, which is also confirmed from Figure \ref{Fig: mass_radius}, indicating super-Chandrasekhar limiting mass. This is similar to the limiting mass proposed by Das and Mukhopadhyay in the presence of magnetic field \cite{2013PhRvL.110g1102D}. 

Let us now discuss about the significance of the NC parameter $\nu_1$. From equation \eqref{rho number density relation}
and \eqref{rho2}, we have
\begin{equation}\label{nu1_new}
\nu_1 = \frac{n_e^2 h^6}{32 \pi^2 \theta^3}.
\end{equation}
Let us now recall the quantum Hall effect, which occurs at high magnetic field and at low temperature in a two dimensional system. In quantum Hall effect, Hall resistivity $\rho_{xy}$ is quantized and it is given by \cite{Prange_Girvin_1990}
\begin{equation}
\rho_{xy} = \frac{h}{e^2}\frac{1}{f},
\end{equation}
where $f$ is the quantization number (also known as the filling factor) and $e$ the charge of electron. In other words, the variation of $\rho_{xy}$ with magnetic field $B$ shows multiple plateau like structure with the values of $\rho_{xy}$ at the plateaus strictly depending on $f$. The values of $B$ at the center of each of these plateaus are given by
\begin{equation}
B = \frac{h n_e}{f e},
\end{equation}
As discussed earlier, in case of NC geometry, $B$ is equivalent of $\theta$, which implies $B=k\theta$ with $k$ being a dimensionful constant. Hence the above expression can be written as
\begin{equation}\label{filling factor}
f = \frac{h n_e}{ek\theta}.
\end{equation}
Now, from equations \eqref{nu1_new} and \eqref{filling factor}, we have
\begin{equation}
n_e\nu_1 = \frac{h^3 e^3 k^3}{32 \pi^2}f^3.
\end{equation}
It is observed that for a fixed $n_e$, $\nu_1\propto f^3$. This implies that $\nu_1$ mimics as the filling factor and hence it represents how much a level is filled in.

\subsection{Subtle issues of classical limit}
From the above discussion, it is important to recall that the NC parameter $\theta$ is equivalent to the magnetic field leading to the Landau quantization. Only difference is that in the conventional Landau quantization, magnetic field is plausibly imposed externally, while in the noncommutativity, it is inherent and spin dependent. The NC parameter is originated from local small-scale curvature owing the gravitational interaction between the electrons and is thus a function of inter-electron separation, hence of the local density. Therefore, $\theta$ decreases from higher density center to lower density  surface, as is evident from equation \eqref{rho2}, leading to the commutative picture as expected at low density. This scenario is similar to the presence of the magnetic field inside stars and of its strength. Therefore, white dwarfs with lower central density should follow Chandrasekhar's EoS only and the corresponding mass-radius relation, unlike what is emerged in other exploration \cite{2019PhLB..79734859P}, which appears to be unphysical.

Moreover, whether the present result also enlightens the progenitors of regular SNeIa, is still a question. The classical Chandrasekhar limit $\sim 1.4M_\odot$ will be obtained in the limit $\theta \to 0$. In other words, if the noncommutativity does not trigger in the density regime of white dwarfs ($\lesssim 10^{10-11}$ g cm$^{-3}$), the original Chandrasekhar mass-limit would remain intact. Of course, the limit $\theta \to 0$ needs to be chosen carefully. In the absence of $\theta$, which is related to density and inter-electron distance, the noncommutativity is not triggered, and conventional Fermi-Dirac statistics works, which along with TOV equations gives the Chandrasekhar mass-limit. On a different scenario, if high density offers noncommutativity leading to a new super-Chandrasekhar mass-limit explaining peculiar over-luminous SNeIa, conventional SNeIa, particularly relatively low luminous SNeIa obeying the pure detonation limit or even combined detonation and deflagration processes \cite{1984ApJ...286..644N,1991A&A...245..114K}, maybe a double-degenerate scenario. Also, at what length scale exactly how coarse-grained/foamy structure emerges in the spacetime is quite uncertain, which is expected to result in a wide range of SNIa luminosities.


\section{Conclusions}\label{Conclusions}

While the idea of noncommutativity of space as well as momentum coordinates is there for quite sometime, there is no direct observational evidence that argues for its indispensable presence. It has been believed for a long time that the effect of noncommutativity is prominent only in the early universe, particularly at the Planck scale, when the density of matter was extremely high. On the other hand, the evidence for at least a dozen of over-luminous peculiar SNeIa argues for highly super-Chandrasekhar progenitor white dwarfs with a limiting mass much larger than the Chandrasekhar-limit. Such a highly super-Chandrasekhar mass-limit is quite evident if the components of position and linear momentum in a plane are assumed to be noncommutative (but on a much weaker scale than those in the Planck regime). This, in turn, affects the white dwarf matter statistically and hence the underlying EoS. This modified EoS leads to a super-Chandrasekhar limiting mass. Earlier attempts to explain super-Chandrasekhar white dwarfs and new limiting mass based on magnetic fields, modifying Einstein's gravity, etc., encounter their respective uncertainties, which may further need to pay attention to repair based on additional physics. In the NC premise, the only required hypothesis is that position and momentum coordinates are related differently at a smaller length scale. Thus, over-luminous peculiar SNeIa suggest possible observational evidences for noncommutativity.

It is generally expected that the NC parameter is controlled by the Planck length $L_P$ and generally the Planck regime. However, in the semi-classical or phenomenological context of a white dwarf, there is a coherent action of all the free electrons when the inter-electron separation could be of the order of or even much less than the Compton wavelength $L_C$ at its center. At this scale, the notion of localizing an electron is lost. At this energy scale, corresponding to this length scale, $e^{-}$--$e^{+}$ pairs are created and, hence, the meaning of the exchange of electrons needs to be modified. Therefore, the statistics of degenerate electrons that plays a role in the Chandrasekhar limit for a typical white dwarf is affected. The two parameters, $L_P$ and $L_C$, control noncommutativity's effectiveness, and the natural length scale for this purpose, as argued by Wigner through position measurement, is $(L_C L_P^2)^{1/3}$, which is higher than the Planck length. Hence, using the system's scale, along with the fundamental scale, the phenomenon needs to be understood. It is to be noted that indeed various literature provides different bounds on the NC parameters (e.g., \cite{2015PhLB..750....6B} and the references therein). It can easily be understood that these bounds depend on the length scale (or equivalently energy scale) of the system. For example, weak equivalence principle provides $\sqrt{\theta}\lesssim 10^{-10}$ eV, whereas high-energy cosmic ray experiment gives a much lighter bound such that $\sqrt{\theta}\lesssim 4 \times 10^7$ eV \cite{2015PhLB..750....6B}. In our calculations, as mentioned in Section \ref{Constraining noncommutativity parameter}, $\theta_D\lesssim287.3$ and, hence, the maximum value of $\sqrt{\theta}$ can be $\sim 8.7 \times 10^6$ eV. This value of $\sqrt{\theta}$ is in accordance with the bound obtained from high-energy cosmic ray, as the length scale of our system is quite higher than the Planck length. In other words, $\sqrt{\theta}=8.7 \times 10^6$ eV is equivalent to the length scale $\sim 10^{-12}$ cm. The parameter $\theta$ (whose origin may be due to gravitational field in semi-classical regime along with very high density of the electron gas) controls the governing statistical mechanics. Similarly, in the context of the quantum Hall effect, the NC parameter is larger depending on the strength of the magnetic field and the nature of electrons. Therefore, a phenomenon can be modeled using the scale provided by the system, which can be different than the Planck scale, whereas a phenomenon, such as the super-Chandrasekhar white dwarf, cannot be used to put limits or bounds on the fundamental NC scale.

In this analysis, based on noncommutativity both in the position and momentum variables, we have arrived at a new mass-limit of the white dwarfs, which is $\sim 2.6 M_\odot$. This is entirely viable with the current observation data, unlike the earlier authors who found the mass-limit to be very high \cite{2019PhLB..79734859P}, which is totally unrealistic. Moreover, we have shown that the NC parameter $\nu_1$ behaves like the filling factor for the Landau levels in the presence of the magnetic field. To obtain the correct mass-limit of the white dwarfs, it is essential to figure out the exact value of $\nu_1$. We have obtained the value of $\nu_1=1/3$ by matching the mass-radius relation for the low density white dwarfs. However, generically electrons in white dwarfs do not necessarily lie in the ground level throughout; hence $\nu$ need not to be always 0, unlike the present simplistic approach. 

In the future, these super-Chandrasekhar white dwarfs may be detected by the space-based gravitational wave detectors like LISA, DECIGO, BBO, etc. \cite{2019MNRAS.490.2692K,2020IAUS..357...79K}, which can enlighten the significance of noncommutativity in white dwarfs.

\section*{Acknowledgement}
BM and TRG acknowledge the hospitality of Max-Planck-Institute for Gravitational Physics, Albert Einstein Institute, Potsdam-Golm, Germany, where the discussion on NC geometry, employed in the present work, was initiated.

\bibliographystyle{ws-ijmpd}
\bibliography{mypaper4}

\begin{thebibliography}{10}

\bibitem{1931ApJ....74...81C}
S.~{Chandrasekhar}, {\em The Astrophysical Journal} {\bf 74} (July 1931)  ~81.

\bibitem{choudhuri_2010}
A.~R. Choudhuri, {\em Astrophysics for Physicists} (Cambridge University Press,
  2010).

\bibitem{2006Natur.443..308H}
D.~A. {Howell}, M.~{Sullivan}, P.~E. {Nugent}, R.~S. {Ellis}, A.~J. {Conley},
  D.~{Le Borgne}, R.~G. {Carlberg}, J.~{Guy}, D.~{Balam}, S.~{Basa},
  D.~{Fouchez}, I.~M. {Hook}, E.~Y. {Hsiao}, J.~D. {Neill}, R.~{Pain}, K.~M.
  {Perrett} and C.~J. {Pritchet}, {\em Nature} {\bf 443} (September 2006) 308,
  \href{http://arxiv.org/abs/astro-ph/0609616}{{\ttfamily astro-ph/0609616}}.

\bibitem{2010ApJ...713.1073S}
R.~A. {Scalzo}, G.~{Aldering}, P.~{Antilogus}, C.~{Aragon}, S.~{Bailey},
  C.~{Baltay}, S.~{Bongard}, C.~{Buton}, M.~{Childress}, N.~{Chotard},
  Y.~{Copin}, H.~K. {Fakhouri}, A.~{Gal-Yam}, E.~{Gangler}, S.~{Hoyer},
  M.~{Kasliwal}, S.~{Loken}, P.~{Nugent}, R.~{Pain}, E.~{P{\'e}contal},
  R.~{Pereira}, S.~{Perlmutter}, D.~{Rabinowitz}, A.~{Rau}, G.~{Rigaudier},
  K.~{Runge}, G.~{Smadja}, C.~{Tao}, R.~C. {Thomas}, B.~{Weaver} and C.~{Wu},
  {\em The Astrophysical Journal} {\bf 713} (April 2010) 1073,
  \href{http://arxiv.org/abs/1003.2217}{{\ttfamily arXiv:1003.2217}}.

\bibitem{2012MPLA...2750084K}
A.~{Kundu} and B.~{Mukhopadhyay}, {\em Modern Physics Letters A} {\bf 27} (May
  2012) 1250084, \href{http://arxiv.org/abs/1204.1463}{{\ttfamily
  arXiv:1204.1463 [astro-ph.SR]}}.

\bibitem{2012PhRvD..86d2001D}
U.~{Das} and B.~{Mukhopadhyay}, {\em Physical Review D} {\bf 86} (Aug 2012)
  042001, \href{http://arxiv.org/abs/1204.1262}{{\ttfamily arXiv:1204.1262
  [astro-ph.HE]}}.

\bibitem{2013PhRvL.110g1102D}
U.~{Das} and B.~{Mukhopadhyay}, {\em Physical Review Letters} {\bf 110} (Feb
  2013)   071102, \href{http://arxiv.org/abs/1301.5965}{{\ttfamily
  arXiv:1301.5965 [astro-ph.SR]}}.

\bibitem{2014JCAP...06..050D}
U.~{Das} and B.~{Mukhopadhyay}, {\em Journal of Cosmology and Astro-Particle
  Physics} {\bf 2014} (Jun 2014)   050,
  \href{http://arxiv.org/abs/1404.7627}{{\ttfamily arXiv:1404.7627
  [astro-ph.SR]}}.

\bibitem{2015JCAP...05..016D}
U.~{Das} and B.~{Mukhopadhyay}, {\em Journal of Cosmology and Astro-Particle
  Physics} {\bf 5} (May 2015)   016,
  \href{http://arxiv.org/abs/1411.5367}{{\ttfamily arXiv:1411.5367
  [astro-ph.SR]}}.

\bibitem{2015MNRAS.454..752S}
S.~{Subramanian} and B.~{Mukhopadhyay}, {\em Monthly Notices of the Royal
  Astronomical Society} {\bf 454} (November 2015) 752,
  \href{http://arxiv.org/abs/1507.01606}{{\ttfamily arXiv:1507.01606
  [astro-ph.SR]}}.

\bibitem{2019MNRAS.490.2692K}
S.~{Kalita} and B.~{Mukhopadhyay}, {\em Monthly Notices of the Royal
  Astronomical Society} {\bf 490} (Dec 2019) 2692,
  \href{http://arxiv.org/abs/1905.02730}{{\ttfamily arXiv:1905.02730
  [astro-ph.HE]}}.

\bibitem{2020IAUS..357...79K}
S.~{Kalita} and B.~{Mukhopadhyay}, {\em IAU Symposium} {\bf 357} (January 2020)
  79, \href{http://arxiv.org/abs/2001.10698}{{\ttfamily arXiv:2001.10698
  [astro-ph.HE]}}.

\bibitem{2015JCAP...05..045D}
U.~{Das} and B.~{Mukhopadhyay}, {\em Journal of Cosmology and Astro-Particle
  Physics} {\bf 5} (May 2015)   045,
  \href{http://arxiv.org/abs/1411.1515}{{\ttfamily arXiv:1411.1515
  [astro-ph.SR]}}.

\bibitem{2017EPJC...77..871C}
G.~A. {Carvalho}, R.~V. {Lobato}, P.~H.~R.~S. {Moraes}, J.~D.~V. {Arba{\~n}il},
  E.~{Otoniel}, R.~M. {Marinho} and M.~{Malheiro}, {\em European Physical
  Journal C} {\bf 77} (December 2017)   871,
  \href{http://arxiv.org/abs/1706.03596}{{\ttfamily arXiv:1706.03596 [gr-qc]}}.

\bibitem{2018JCAP...09..007K}
S.~{Kalita} and B.~{Mukhopadhyay}, {\em Journal of Cosmology and Astro-Particle
  Physics} {\bf 2018} (Sep 2018)   007,
  \href{http://arxiv.org/abs/1805.12550}{{\ttfamily arXiv:1805.12550 [gr-qc]}}.

\bibitem{2016PhRvD..93j4046B}
O.~{Bertolami} and H.~{Mariji}, {\em Physical Review D} {\bf 93} (May 2016)
  104046, \href{http://arxiv.org/abs/1603.09282}{{\ttfamily arXiv:1603.09282
  [astro-ph.SR]}}.

\bibitem{2015NuPhA.937...17B}
V.~B. {Belyaev}, P.~{Ricci}, F.~{{\v{S}}imkovic}, J.~{Adam}, M.~{Tater} and
  E.~{Truhl{\'\i}k}, {\em Nuclear Physics A} {\bf 937} (May 2015) 17,
  \href{http://arxiv.org/abs/1212.3155}{{\ttfamily arXiv:1212.3155 [nucl-th]}}.

\bibitem{2014PhRvD..89j4043L}
H.~{Liu}, X.~{Zhang} and D.~{Wen}, {\em Physical Review D} {\bf 89} (May 2014)
   104043, \href{http://arxiv.org/abs/1405.3774}{{\ttfamily arXiv:1405.3774
  [gr-qc]}}.

\bibitem{2018JCAP...09..015O}
Y.~C. {Ong}, {\em Journal of Cosmology and Astro-Particle Physics} {\bf 9}
  (September 2018)   015, \href{http://arxiv.org/abs/1804.05176}{{\ttfamily
  arXiv:1804.05176 [gr-qc]}}.

\bibitem{2019PhLB..79734859P}
S.~K. {Pal} and P.~{Nandi}, {\em Physics Letters B} {\bf 797} (Oct 2019)
  134859, \href{http://arxiv.org/abs/1908.11206}{{\ttfamily arXiv:1908.11206
  [gr-qc]}}.

\bibitem{2004ApJ...615..444S}
H.~{Saio} and K.~{Nomoto}, {\em The Astrophysical Journal} {\bf 615} (November
  2004) 444, \href{http://arxiv.org/abs/astro-ph/0401141}{{\ttfamily
  arXiv:astro-ph/0401141 [astro-ph]}}.

\bibitem{2006MNRAS.373..263M}
R.~G. {Martin}, C.~A. {Tout} and P.~{Lesaffre}, {\em Monthly Notices of the
  Royal Astronomical Society} {\bf 373} (November 2006) 263,
  \href{http://arxiv.org/abs/astro-ph/0609192}{{\ttfamily
  arXiv:astro-ph/0609192 [astro-ph]}}.

\bibitem{2009ApJ...702..686C}
W.-C. {Chen} and X.-D. {Li}, {\em The Astrophysical Journal} {\bf 702}
  (September 2009) 686, \href{http://arxiv.org/abs/0907.0057}{{\ttfamily
  arXiv:0907.0057 [astro-ph.HE]}}.

\bibitem{2019MNRAS.483..263W}
C.~{Wu}, B.~{Wang} and D.~{Liu}, {\em Monthly Notices of the Royal Astronomical
  Society} {\bf 483} (February 2019) 263,
  \href{http://arxiv.org/abs/1811.08638}{{\ttfamily arXiv:1811.08638
  [astro-ph.SR]}}.

\bibitem{2010Natur.463...61P}
R.~{Pakmor}, M.~{Kromer}, F.~K. {R{\"o}pke}, S.~A. {Sim}, A.~J. {Ruiter} and
  W.~{Hillebrandt}, {\em Nature} {\bf 463} (January 2010) 61,
  \href{http://arxiv.org/abs/0911.0926}{{\ttfamily arXiv:0911.0926
  [astro-ph.HE]}}.

\bibitem{2014NewAR..62...15R}
P.~{Ruiz-Lapuente}, {\em New Astronomy Reviews} {\bf 62} (October 2014) 15,
  \href{http://arxiv.org/abs/1403.4087}{{\ttfamily arXiv:1403.4087
  [astro-ph.HE]}}.

\bibitem{2018PhR...736....1L}
M.~{Livio} and P.~{Mazzali}, {\em Physics Reports} {\bf 736} (March 2018) 1,
  \href{http://arxiv.org/abs/1802.03125}{{\ttfamily arXiv:1802.03125
  [astro-ph.SR]}}.

\bibitem{2019arXiv191201550S}
N.~{Soker}, {\em arXiv e-prints}  (December 2019)   arXiv:1912.01550,
  \href{http://arxiv.org/abs/1912.01550}{{\ttfamily arXiv:1912.01550
  [astro-ph.HE]}}.

\bibitem{1992CQGra...9...69M}
J.~{Madore}, {\em Classical and Quantum Gravity} {\bf 9} (Jan 1992) 69.

\bibitem{2003JHEP...08..057L}
F.~{Lizzi}, P.~{Vitale} and A.~{Zampini}, {\em Journal of High Energy Physics}
  {\bf 2003} (Aug 2003)   057,
  \href{http://arxiv.org/abs/hep-th/0306247}{{\ttfamily arXiv:hep-th/0306247
  [hep-th]}}.

\bibitem{2014JPhA...47R5203C}
N.~{Chandra}, H.~W. {Groenewald}, J.~N. {Kriel}, F.~G. {Scholtz} and
  S.~{Vaidya}, {\em Journal of Physics A Mathematical General} {\bf 47} (Nov
  2014)   445203, \href{http://arxiv.org/abs/1407.5857}{{\ttfamily
  arXiv:1407.5857 [hep-th]}}.

\bibitem{2015JPhA...48C5401A}
S.~{Andronache} and H.~C. {Steinacker}, {\em Journal of Physics A Mathematical
  General} {\bf 48} (Jul 2015)   295401,
  \href{http://arxiv.org/abs/1503.03625}{{\ttfamily arXiv:1503.03625
  [hep-th]}}.

\bibitem{2015PhRvD..92l5013S}
F.~G. {Scholtz}, J.~N. {Kriel} and H.~W. {Groenewald}, {\em Physical Review D}
  {\bf 92} (Dec 2015)   125013,
  \href{http://arxiv.org/abs/1508.05799}{{\ttfamily arXiv:1508.05799
  [hep-th]}}.

\bibitem{2006LNP...698...97N}
V.~P. {Nair}, {Noncommutative Mechanics, Landau Levels, Twistors, and
  Yang-Mills Amplitudes}, in {\em Lecture Notes in Physics, Berlin Springer
  Verlag\/},  ed. S.~{Bellucci} 2006, p.~97.

\bibitem{2017EPJC...77..577K}
R.~{Kumar} and S.~G. {Ghosh}, {\em European Physical Journal C} {\bf 77} (Sep
  2017)   577, \href{http://arxiv.org/abs/1703.10479}{{\ttfamily
  arXiv:1703.10479 [gr-qc]}}.

\bibitem{2018EPJP..133..421F}
S.~A. {Franchino-Vi{\~n}as} and P.~{Pisani}, {\em European Physical Journal
  Plus} {\bf 133} (Oct 2018)   421,
  \href{http://arxiv.org/abs/1803.05821}{{\ttfamily arXiv:1803.05821
  [hep-th]}}.

\bibitem{1999JHEP...09..032S}
N.~{Seiberg} and E.~{Witten}, {\em Journal of High Energy Physics} {\bf 1999}
  (Sep 1999)   032, \href{http://arxiv.org/abs/hep-th/9908142}{{\ttfamily
  arXiv:hep-th/9908142 [hep-th]}}.

\bibitem{2000IJMPA..15.4301A}
G.~{Amelino-Camelia} and S.~{Majid}, {\em International Journal of Modern
  Physics A} {\bf 15} (Jan 2000) 4301,
  \href{http://arxiv.org/abs/hep-th/9907110}{{\ttfamily arXiv:hep-th/9907110
  [hep-th]}}.

\bibitem{2001PhLB..510..255A}
G.~{Amelino-Camelia}, {\em Physics Letters B} {\bf 510} (Jun 2001) 255,
  \href{http://arxiv.org/abs/hep-th/0012238}{{\ttfamily arXiv:hep-th/0012238
  [hep-th]}}.

\bibitem{2002PhRvL..88s0403M}
J.~{Magueijo} and L.~{Smolin}, {\em Physical Review Letters} {\bf 88} (May
  2002)   190403, \href{http://arxiv.org/abs/hep-th/0112090}{{\ttfamily
  arXiv:hep-th/0112090 [hep-th]}}.

\bibitem{2002Natur.418...34A}
G.~{Amelino-Camelia}, {\em Nature} {\bf 418} (Jul 2002) 34,
  \href{http://arxiv.org/abs/gr-qc/0207049}{{\ttfamily arXiv:gr-qc/0207049
  [gr-qc]}}.

\bibitem{2012PhRvL.109r1602S}
D.~T. {Son} and N.~{Yamamoto}, {\em Physical Review Letters} {\bf 109} (Nov
  2012)   181602, \href{http://arxiv.org/abs/1203.2697}{{\ttfamily
  arXiv:1203.2697 [cond-mat.mes-hall]}}.

\bibitem{2006PhLB..632..547N}
P.~{Nicolini}, A.~{Smailagic} and E.~{Spallucci}, {\em Physics Letters B} {\bf
  632} (Jan 2006) 547, \href{http://arxiv.org/abs/gr-qc/0510112}{{\ttfamily
  arXiv:gr-qc/0510112 [gr-qc]}}.

\bibitem{2007PhRvD..75d5009B}
A.~P. {Balachandran}, T.~R. {Govindarajan}, G.~{Mangano}, A.~{Pinzul}, B.~A.
  {Qureshi} and S.~{Vaidya}, {\em Physical Review D} {\bf 75} (Feb 2007)
  045009, \href{http://arxiv.org/abs/hep-th/0608179}{{\ttfamily
  arXiv:hep-th/0608179 [hep-th]}}.

\bibitem{2006JPhA...39.9557C}
B.~{Chakraborty}, S.~{Gangopadhyay}, A.~{Ghosh Hazra} and F.~G. {Scholtz}, {\em
  Journal of Physics A Mathematical General} {\bf 39} (Jul 2006) 9557,
  \href{http://arxiv.org/abs/hep-th/0601121}{{\ttfamily arXiv:hep-th/0601121
  [hep-th]}}.

\bibitem{2013IJMPD..2242004D}
U.~{Das} and B.~{Mukhopadhyay}, {\em International Journal of Modern Physics D}
  {\bf 22} (Sep 2013)   1342004,
  \href{http://arxiv.org/abs/1305.3987}{{\ttfamily arXiv:1305.3987
  [astro-ph.HE]}}.

\bibitem{2018MNRAS.480.4505J}
F.~M. {Jim{\'e}nez-Esteban}, S.~{Torres}, A.~{Rebassa-Mansergas},
  G.~{Skorobogatov}, E.~{Solano}, C.~{Cantero} and C.~{Rodrigo}, {\em Monthly
  Notices of the Royal Astronomical Society} {\bf 480} (Nov 2018) 4505,
  \href{http://arxiv.org/abs/1807.02559}{{\ttfamily arXiv:1807.02559
  [astro-ph.SR]}}.

\bibitem{1949PCPS...45...99M}
J.~E. {Moyal} and M.~S. {Bartlett}, {\em Proceedings of the Cambridge
  Philosophical Society} {\bf 45} (Jan 1949)  ~99.

\bibitem{1946Phy....12..405G}
H.~J. {Groenewold}, {\em Physica} {\bf 12} (Oct 1946) 405.

\bibitem{2011PhRvD..83h4003C}
W.~A. {Christiansen}, Y.~J. {Ng}, D.~J.~E. {Floyd} and E.~S. {Perlman}, {\em
  Physical Review D} {\bf 83} (Apr 2011)   084003,
  \href{http://arxiv.org/abs/0912.0535}{{\ttfamily arXiv:0912.0535
  [astro-ph.CO]}}.

\bibitem{1958PhRv..109..571S}
H.~{Salecker} and E.~P. {Wigner}, {\em Physical Review} {\bf 109} (Jan 1958)
  571.

\bibitem{2003MPLA...18.1073N}
Y.~J. {Ng}, {\em Modern Physics Letters A} {\bf 18} (Jan 2003) 1073,
  \href{http://arxiv.org/abs/gr-qc/0305019}{{\ttfamily arXiv:gr-qc/0305019
  [gr-qc]}}.

\bibitem{2019Entrp..21.1035N}
Y.~J. {Ng}, {\em Entropy} {\bf 21} (Oct 2019)   1035.

\bibitem{2009igr..book.....R}
L.~{Ryder}, {\em {Introduction to General Relativity}} (Cambridge University
  Press., 2009).

\bibitem{1964ApJ...140..434T}
R.~F. {Tooper}, {\em The Astrophysical Journal} {\bf 140} (Aug 1964)   434.

\bibitem{1973ApJ...183..637B}
S.~A. {Bludman}, {\em The Astrophysical Journal} {\bf 183} (Jul 1973) 637.

\bibitem{Prange_Girvin_1990}
R.~E. {Prange} and S.~M. {Girvin}, {\em The Quantum Hall Effect}
  (Springer-Verlag New York, 1990).

\bibitem{1984ApJ...286..644N}
K.~{Nomoto}, F.~K. {Thielemann} and K.~{Yokoi}, {\em The Astrophysical Journal}
  {\bf 286} (Nov 1984) 644.

\bibitem{1991A&A...245..114K}
A.~M. {Khokhlov}, {\em Astronomy \& Astrophysics} {\bf 245} (May 1991) 114.

\bibitem{2015PhLB..750....6B}
O.~{Bertolami} and P.~{Leal}, {\em Physics Letters B} {\bf 750} (November 2015)
  6, \href{http://arxiv.org/abs/1507.07722}{{\ttfamily arXiv:1507.07722
  [gr-qc]}}.

\end{thebibliography}

\end{document}